\documentclass[12pt]{article}

\usepackage{epsf}
\usepackage{graphicx}
\usepackage{amssymb}

\def\lsim{\mathrel{\raise.3ex\hbox{$<$\kern-.75em\lower1ex\hbox{$\sim$}}}} 
\def\gsim{\mathrel{\raise.3ex\hbox{$>$\kern-.75em\lower1ex\hbox{$\sim$}}}} 
\newcommand\apj{{ApJ}}%
\def\simlt{\ \raise -2.truept\hbox{\rlap{\hbox{$\sim$}}\raise5.truept   %
\hbox{$<$}\ }}
\def\simgt{\ \raise -2.truept\hbox{\rlap{\hbox{$\sim$}}\raise5.truept   %
\hbox{$>$}\ }}                                                          %

\def\be{\begin{equation}}
\def\ee{\end{equation}}
\def\newline{\hfil\break}

\def\ergcm2s{{erg~cm$^{-2}$s$^{-1}$~}}

\def\la{\mathrel{\hbox{\rlap{\hbox{\lower4pt\hbox{$\sim$}}}\hbox{$<$}}}}
\def\ga{\mathrel{\hbox{\rlap{\hbox{\lower4pt\hbox{$\sim$}}}\hbox{$>$}}}}

\def\mug{$\mu$G ~}
\def\be{\begin{equation}}
\def\ee{\end{equation}}
\def\cm3{cm$^{-3}$}
\def\es{{\rm 1ES0657-556}~}

\def\sigva{\langle\sigma v\rangle}

\begin{document}

\pagenumbering{roman}
\pagenumbering{arabic}

\begin{center}
{\LARGE \mbox{Multi-wavelength Searches for Particle Dark Matter}}\footnote{\it This contribution appeared as chapter 27, pp. 547-564, of  `Particle Dark Matter: Observations, Models and Searches'
edited by Gianfranco Bertone,
Copyright 2010 Cambridge University Press.
Hardback  ISBN  9780521763684, 
http://cambridge.org/us/catalogue/catalogue.asp?isbn=9780521763684}
\end{center}

\vspace{0.2cm}

\begin{center}
{\large Stefano Profumo$^{a}$, Piero Ullio$^{b,c}$}\\[0.3cm]

{\it \small
 $^a$ Department of Physics and Santa Cruz Institute for Particle Physics,\\ University of California, Santa Cruz, CA 95064, USA \\ \mbox{$^b$ Scuola Internazionale Superiore di Studi Avanzati, Via Beirut 2-4, I-34014 Trieste, Italy}\\ \mbox{$^c$ Istituto Nazionale di Fisica Nucleare, Sezione di Trieste, I-34014 Trieste, Italy}
}
\end{center}

\vskip 0.4cm

\abstract{\noindent If dark matter particles pair annihilate into stable Standard Model particles, a population of energetic, non-thermal electrons and positrons is necessarily produced. The secondary radiation resulting from the energy losses of this non-standard population (including synchrotron and inverse Compton up-scattering of background radiation photons) in turn produces a peculiar multi-wavelength spectrum, extending from radio to gamma-ray frequencies. We give here an overview of multi-wavelength searches for dark matter, including the expected injection spectrum and production rate of electrons and positrons from dark matter annihilation, the computation of the effects of propagation and energy losses and the actual multi-wavelength emission. We then outline the application of this general framework to the case of galaxy clusters (specifically Coma, 1E 0657-56 - the so-called Bullet cluster - and Ophiuchus) and of dwarf spheroidal galaxies (including Draco, Fornax, Ursa Minor and Carina). We also review the application of multi-wavelength search strategies to our own Milky Way, and more specifically to the Galactic center environment, to dark galactic mini-halos and to the so-called WMAP haze and other radio data. We argue that multi-wavelength observations will complement gamma-ray observations as probes of particle dark matter, since the expected luminosities at different frequencies are generically comparable. The indirect search for dark matter with astronomical observations at various frequencies is therefore a crucial and indispensable element in the quest for the fundamental nature of dark matter.}


\section{Introduction}

The class of Weakly interacting massive particles (WIMPs) is the leading category of particle dark matter candidates: On the one hand, the mechanism of thermal freeze-out of a stable WIMP $\chi$ leads to a non-relativistic relic population whose relative matter density can be approximated as $\Omega_\chi h^2\approx3\times 10^{-27}\ {\rm cm}^3{\rm s}^{-1}/\langle\sigma_A v\rangle$, where $h$ is the Hubble constant in units of 100 ${\rm km}\ {\rm s}^{-1}\ {\rm Mpc}^{-1}$, $\Omega_\chi$ is the ratio of the $\chi$ density over the critical density, and $\langle\sigma_A v\rangle$ is the thermally averaged $\chi$ pair-annihilation cross section. If the new physics connected to the WIMP is at the electro-weak scale, $m_{\rm EW}$, one can estimate $\langle\sigma_A v\rangle\approx \alpha_{\rm EW}^2/m^2_{\rm EW}\approx 10^{-25}\ {\rm cm}^3{\rm s}^{-1}$, giving a thermal leftover $\chi$ population that lies in the same ballpark as the dark matter abundance inferred from cosmic microwave background anisotropies, large scale structure and other astronomical observations \cite{Komatsu:2008hk}; On the other hand, numerous motivated particle physics extensions to the Standard Model encompass a stable WIMP, including supersymmetry \cite{Jungman:1995df} and universal extra-dimensions (UED) \cite{Hooper:2007qk}, in virtue of unbroken discrete symmetries ($R$-parity in the case of supersymmetry, and Kaluza-Klein parity in the case of UED).

Since WIMPs were once kept in thermal equilibrium by pair annihilations into Standard Model particles and inverse WIMP pair production processes, even in today's cold Universe, occasionally, WIMPs can pair annihilate, giving rise to energetic, stable ``ordinary'' Standard Model particles. The rate of WIMP pair annihilation depends on the number density of dark matter in a given location and on the pair-annihilation rate $\langle\sigma_A v\rangle$. The spectrum and particle content of the final state products of a pair-annihilation event are, instead, determined once the microscopic details of the specific WIMP model are specified. Given a WIMP $\chi$, the production rate of a final state Standard Model particle species $f=\gamma,\ e^\pm, \bar p,\ \nu,\ldots$, emitted in a pair-annihilation event in a location $\vec x$ with dark matter density $\rho_{\rm DM}(\vec x)$, reads
\begin{equation}\label{eq:indirect}
Q(p,\vec x) \equiv \frac{\rho^2_{\rm DM}(\vec x)}{2m_\chi}\langle\sigma_A v\rangle\frac{{\rm d}N_f}
{{\rm d}p}.
\end{equation}
In the formula above, ${\rm d}N_f$ is the number of $f$ particles with modulus of the momenta between $p$ and $p+{\rm d}p$ resulting from the WIMP pair-annihilation event. Numerous studies have investigated the {\em indirect} detection of WIMPs through signatures in the spectrum of cosmic-ray positrons and antiprotons, or through the emission of energetic gamma rays. With the recent successful launch of the space-based antimatter detector PAMELA \cite{Picozza:2006nm} and of the Fermi gamma-ray space telescopes \cite{Baltz:2008wd}, dark matter indirect detection faces new opportunities, with a forthcoming dramatic increase in the quality and quantity of available data.

The pair-annihilation of WIMPs yields gamma rays from the two photon decay of neutral pions, and, in the high energy end of the spectrum, from internal bremsstrahlung from charged particle final states. 
Sub-dominant one-loop annihilation modes give rise to the additional emission of monochromatic photons: the direct pair-annihilation into two photons, for instance, yields two gamma rays with $E_\gamma=m_\chi$. Photons from prompt emission are generally in the gamma-ray band, where 
absorption is negligible, hence the induced fluxes or intensities can be straightforwardly derived summing contributions along the line of sight. 

Concerning lower frequencies, a conspicuous non-thermal population of energetic electrons and positrons results from the decays of charged pions produced by the hadronization of strongly interacting particles in the final state, as well as from the decays of gauge bosons, Higgs bosons and charged leptons. This non-thermal $e^\pm$ population looses energy and produces secondary radiation through several processes: synchrotron in the presence of magnetic fields, inverse Compton scattering off starlight and cosmic microwave background photons and bremsstrahlung in the presence of ionized gas. This radiation can actually cover the whole electromagnetic spectrum between the radio band to
the gamma-ray band.

The computation of the multiwavelength emissions from WIMP-induced energetic $e^\pm$ involves several steps: (1) the assessment of the spectrum and production rate of electrons and positrons from dark matter pair annihilations (that follows from Eq.~(\ref{eq:indirect}) with $f=e^\pm$); (2) the computation of the effects of propagation and energy losses, possibly leading to a steady state $e^\pm$ configuration; (3) the computation of the actual emissions from the mentioned equilibrium configuration;
(4) the evaluation of eventual absorption of the emitted radiation along the line of sight to derive fluxes and intensities for a local observer. We describe items (1) and (2) in the next section \ref{sec:epemdistrib}, while we devote sec.~\ref{sec:emissions} to the calculation of the secondary emission from dark matter annihilation induced $e^\pm$. Sec.~\ref{sec:clusters} then applies the general outlined framework to the case of galaxy clusters, specifically Coma (sec.~\ref{sec:coma}), 1E 0657-56 (the so-called Bullet cluster, \ref{sec:bul}) and Ophiuchus (\ref{sec:oph}). The following sec.~\ref{sec:dwarfs} studies the other extreme of galactic scales: dwarf galaxies. In particular, we focus on the case of the dwarf spheroidal (dSph) galaxy Draco in \ref{sec:draco} and on constraints on dark matter particle models from X-ray observations of dSph galaxies Fornax, Ursa Minor and Carina in \ref{sec:xraydwarf}. Sec.~\ref{sec:mw} specializes to the case of our own Milky Way, and more specifically on the Galactic center (\ref{sec:gc}), on dark galactic mini-halos (\ref{sec:minihalos}) and on the dark matter annihilation origin of the so-called WMAP haze (\ref{sec:haze}). Finally, sec.~\ref{sec:radio} addresses radio constraints from cosmological dark matter annihilation at all redshifts, and sec.~\ref{sec:concl} concludes and provides an overview.

\subsection{The electron-positron equilibrium number densities}\label{sec:epemdistrib}

Except for WIMPs which annihilate mainly into light leptons, the bulk of electrons and positrons 
from WIMP annihilations are secondaries from intermediate steps in which charged pions 
are produced and decay along the chain: $\pi^{\pm}\to \mu^{\pm} \nu_{\mu}(\bar{\nu}_{\mu})$, with
$\mu^{\pm}\to e^{\pm} + \bar{\nu}_{\mu}(\nu_{\mu}) + \nu_e (\bar{\nu}_e)$. These are twin 
processes with respect to the production of neutral pions and their $\pi^0 \to \gamma \gamma$
decay, hence at the level of the production rates, the yield into electron and positrons in 
Eq.~\ref{eq:indirect} is at the same level of the gamma-ray yield and with analogous spectral 
shape. In principle one should take into account other stable yields, such as protons,  which can be
confined in the astrophysical environment in which WIMP annihilation is considered, and
subsequently give rise to further secondaries (such as pions in proton collisions with the
ambient interstellar material); their multiplicity in WIMP annihilation is however much smaller 
than for electrons and positrons and, therefore, they can be safely neglected.

After emission, electrons and positrons go through a random walk in the turbulent and regular components of ambient magnetic field, loosing energy by radiative emissions, as well as
loosing and/or gaining energy in the convective or advective effects associated to plasma outflows
or inflows. The sum of these effects are usually described in terms of a transport equation 
of the form:
\begin{eqnarray}
{\partial n_e (\vec x,p,t) \over \partial t} 
&= &
  Q_e(\vec r, p, t)                                             
   + \vec\nabla \cdot ( D_{xx}\vec\nabla n_e)
   + {\partial\over\partial p}\, p^2 D_{pp} {\partial\over\partial p}\, {1\over p^2}\, n_e   \nonumber  \\
   && + {\partial\over\partial p} \left(\dot{p} \,n_e \right)
   - \vec\nabla \cdot (\vec V n_e) + {\partial\over\partial p} 
   \left[{p\over 3} \, (\vec\nabla \cdot \vec V )n_e\right] 
\label{eq:prop_n}
\end{eqnarray}
where $n_e(\vec r, p, t)$ is electron/positron number density per unit particle momentum $p$,
$D_{xx}$ is the spatial diffusion coefficient, diffusive re-acceleration is described as diffusion 
in momentum space and is determined by the coefficient $D_{pp}$, 
$\dot{p}\equiv dp/dt$ is the momentum gain or loss rate, and $\vec V$ is the convection 
or advection velocity. Except for when considering the effect of individual dark matter 
substructures orbiting within a given astrophysical object 
the WIMP source function can be regarded as time independent, and the emission rate is 
large enough to reach equilibrium, so that Eq.~\ref{eq:prop_n} can be solved in the steady 
state approximation, i.e. setting the left-hand-side of  Eq.~\ref{eq:prop_n} to zero. Simplifying 
ans\"azte on the morphology of the source (namely spherical symmetry when considering 
galaxy clusters, dwarf satellites and the galactic center region, axial symmetry when considering
the whole Milky Way) and on the spatial dependence of the parameters in the model
are usually adopted;  another useful approximation is to consider
the limit of free-escape at the boundary of the diffusion region, i.e. impose $n_e=0$ outside
the region where it is estimated that the propagation equation can be applied. We discuss 
briefly below how relevant of the different terms in the transport equation are for astrophysical
sources we will consider.

\subsubsection{Energy loss term}

The radiative losses affecting the $e^+-e^-$ propagation are  inverse Compton scattering on CMB and starlight, synchrotron emission, bremsstrahlung, ionization and Coulomb scattering.
The term $\dot{p}$ in Eq.~\ref{eq:prop_n}  can be written as:
\begin{eqnarray}
  \dot{p} & = &  \dot{p}_{IC}(p) + \dot{p}_{syn}(p) +  \dot{p}_{brem}(p) + \dot{p}_{ion}(p) +
  \dot{p}_{Coul}(p)  \nonumber \\
  &\simeq & \dot{p}_{IC}(p) + \dot{p}_{syn}(p)  \nonumber \\
  &\simeq & b_{IC}^0  \frac{U_{ph}}{\rm{1\;eV\,cm}^{-3}} 
  \left(\frac{p}{\rm{1\;GeV}}\right)^{2} + b_{syn}^0 \left(\frac{B}{1\;\mu\rm{G}}\right)^{2}
  B_\mu^2 \left(\frac{p}{\rm{1\;GeV}}\right)^{2}\,
\end{eqnarray}
where we have indicated that in all cases of our interest, the relativistic regime,  
either inverse Compton or synchrotron
dominate, and the coefficients $b_{IC}^0 \simeq 0.76$ and $b_{syn}^0 \simeq 0.025$ are found,
respectively, for a background photon energy density $U_{ph}$ of  $\rm{1\;eV\,cm}^{-3}$
(the CMB energy density is about $0.25\,\rm{eV\,cm}^{-3}$, while the starlight energy density
in the Milky Way is about $0.6\,\rm{eV\,cm}^{-3}$ locally and $8\,\rm{eV\,cm}^{-3}$ in the galactic 
center region~\cite{Porter:2008ve}) and a radiating magnetic field $B$ of a $1\;\mu\rm{G}$ strength
(say, within one order of magnitude of what is expected in galaxy clusters, dwarf galaxies
or the local portion of the Milky Way, while in the Galactic center region magnetic fields should be 
much stronger, at the $1\,\rm{G}$ level or even higher).
 
\subsubsection{Spatial diffusion and re-acceleration}

There is no definite prediction for the diffusion coefficient $D_{xx}(p)$. Its spectral shape $\alpha$
and normalization depend on unknown variables describing turbulence, namely the amplitude of the 
random magnetic field and the scale and the spectrum of the turbulence. E.g., one finds that 
in case of Bohm diffusion $\alpha =1$, while for  Kolmogorov and Kraichnan diffusion, the scaling 
in energy is, respectively, $\alpha=1/3$ and $\alpha=1/2$. On top of this, there is the uncertainty 
on whether and how to include the re-acceleration term $D_{pp}(p)$, which is in general predicted
to be related to  $D_{xx}(p)$. When considering the case for diffusion of cosmic rays within the 
Milky Way one can rely on the comparison between the model and a rather vast and variegated  
set of observables, such as the ratio of local measurements of abundances of secondary versus 
primary cosmic-ray species, and global maps of photon emissivities in different energy bands.
A nearly self-consistent picture (see some more discussion below) can be obtained for, say,  
a diffusion coefficient at 1~GeV of the order of $\sim 10^{28} \rm{cm}^2\rm{s}^{-1}$, 
$\alpha\sim 0.6$ and no re-acceleration, or  $\alpha=1/3$ and including re-acceleration
(see, e.g.~\cite{Strong:2007nh}). 

Assuming, for simplicity, that re-acceleration can be neglected, to have a feeling on whether 
diffusion is relevant,  we can guess, on dimensional grounds, that the diffusion length scale
should be of the order of the square root of the diffusion coefficient times the time scale 
relevant in the transport equation,  i.e. $\lambda_{xx} \sim \sqrt{D_{xx} \tau_{loss}}$, with
$\tau_{loss}$ the time scale for the energy loss associated to radiative processes, namely
$\tau_{loss} \sim p/\dot p$. This guess gives a result that is not too far from the diffusion length scale 
one obtains in one of the limits in which the transport equation can be solved analytically, namely
for spatially constant diffusion coefficient and energy loss term, and neglecting re-acceleration and  convection~\cite{Baltz:1998xv}: $\lambda_{xx}(p) = 2\sqrt{v(p)-v(p')}$, 
where we introduced the variable $v$, in place of the momentum $p$ of the radiating 
electron/positron,  through the double change of variables:
\begin{equation}
v=\int_{u_{\rm min}}^u d\tilde{u} D(\tilde{u})  \;\;\;\; {\rm and} \;\;\;\; 
u = \int_p^{p_{\rm max}} \frac{d\tilde{p}}{\dot{p}(\tilde{p})}
\end{equation}
and $p'$ is the positron/electron momentum at emission.
%
\begin{figure}[!t]
\begin{center}
\hspace*{-0.5cm}\includegraphics[width=7.8cm]{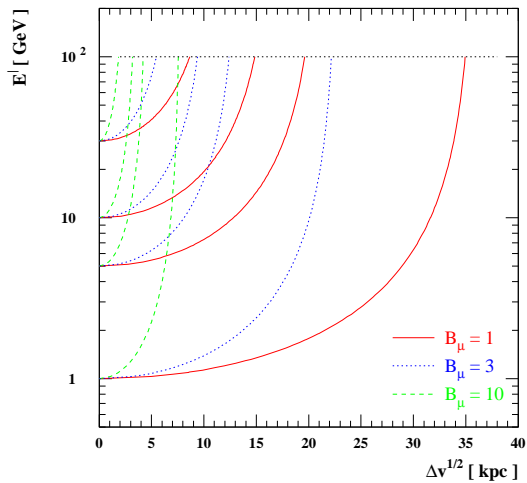}\\
\end{center}
\caption{The figure shows the distance $(\Delta v)^{1/2}$ which, on average, a
electron covers while losing energy from its energy at emission $E^{\prime}$ and the
energy when it interacts $E$, for a few values of $E$: 30~GeV, 10~GeV, 5~GeV and 1~GeV,
and for a few values of the magnetic field (in $\mu$G); we are focusing on a WIMP of mass
100~GeV, hence cutting $E^{\prime} < 100$~GeV.}
\label{fig:deltav}
\end{figure}
%
As a sample case, we plot in Fig.~\ref{fig:deltav}  the square root of $\Delta v \equiv v(p) - v(p')$, for a
few values of radiating and emission energies, for a few sample values of the ambient 
magnetic field in units of $\mu$G~$B_\mu$ and with diffusion coefficient within an
extrapolation which should apply to galaxy clusters (see e.g. 
\cite{Colafrancesco:1998us,Blasi:1999aj}). Length scales of few (tens of) $kpc$ should be 
compared to the typically scales over which the dark matter density profiles varies or to
the length scale corresponding, for a given object distance, to the angular resolution of the 
detector searching for the induced signal. In galaxy clusters the dark matter halo scale 
parameter is generally much larger, and diffusion turns out to be relevant only for dark matter
profiles which are singular towards their center, on angular portions on the sky which are 
however small compared detector resolutions; it follows that spatial diffusion is a marginal
effect in galaxy clusters, and can be safely neglected for most practical purposes in this case
~\cite{Colafrancesco:2005ji}. In the limit of negligible spatial diffusion (as well as negligible re-acceleration and  convection), the equilibrium number density $n_e$ reduces to:
\begin{equation}
n_e(\vec x, p) =  \frac{1}{\dot p(p)} \int_p^{M_\chi}
dp'  \; Q_e(\vec x ,p')\;. \label{eq:nodiff}
\end{equation}
On the other hand, in case of Milky Way dwarf satellites (whose size is few kpc), and for estimates 
of the induced signals in the local neighborhood or even extended portions of the Milky Way 
(with Milky Way  diffusion region extending only few kpc above the Galactic plane, up to, possibly, 10~kpc or so~\cite{Strong:2001gh}), spatial propagation of electrons and positrons is an important 
effect and should be appropriately modeled (see, e.g., \cite{Baltz:1998xv,Colafrancesco:2005ji}).

\subsubsection{Convection or advection effects}

In the Galactic center region the energy loss time--scale decreases dramatically, since
synchrotron radiation losses are busted by the very large magnetic field; still, one cannot
simply refer to Eq.~\ref{eq:nodiff}. The last two terms on the right hand side of 
Eq.~\ref{eq:prop_n} need to be included to model, in the innermost part of the Galaxy,
the accretion flow of gas onto the central black hole~\cite{Melia:1992,Aloisio:2004hy}.
In the limit of radial free fall $\vec V$ is oriented towards the black hole and scales like
$-c\sqrt{{R_{BH}}/{r}}$ (with $R_{BH}$  the gravitational radius of the black hole, and $r$
the radial distance) up to the accretion radius which is about 
$R_{acc}\sim 0.04$~pc~\cite{Melia:1992}. A particle 
propagating in such accretion flow gains momentum since it feels an adiabatic compression 
in the black hole direction. 
%
\begin{figure}[!t]
\begin{center}
\hspace*{-0.5cm}\includegraphics[width=7.8cm]{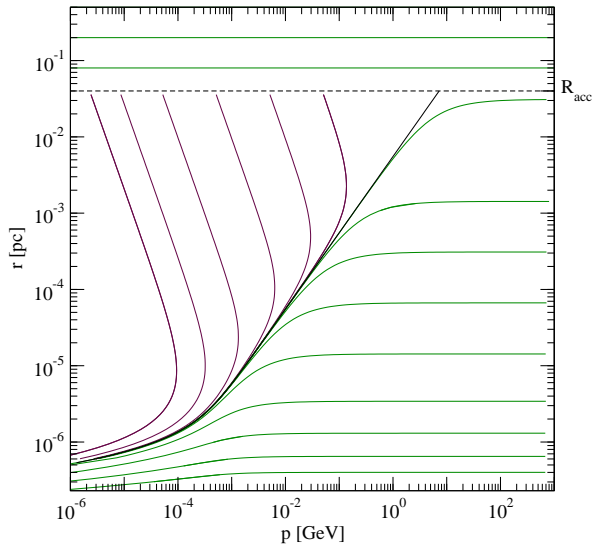}\\
\end{center}
\caption{Electron/positron trajectories in the plane radius versus momentum in case of magnetic 
field for the Galactic center region satisfying the equipartition condition. Far from the turning points, synchrotron loss is dominant in green trajectories, while adiabatic heating takes over in violet trajectories. The black solid line represents the curve along which the $e^+ -e^-$ accumulate since the two effects balance each other. The dotted line is the accretion radius $R_{acc}=0.04\,pc$, where advection is assumed to stop.}
\label{fig:traj}
\end{figure}
%
In Fig.~\ref{fig:traj} we show the interplay between synchrotron loss and adiabatic heating, 
plotting electron/positron trajectories in the plane radial distance versus particle momentum,
in the limit of equipartition for the magnetic field in the central region of the Galaxy, i.e.
assuming he magnetic energy completely balancing the kinetic pressure~ \cite{Melia:1992,Aloisio:2004hy}. The synchrotron loss dominates at high energies, while the advection gain takes over at low energies; electrons accumulate on the trajectory separating the two regimes (black curve in the figure). Since approaching the BH, the scaling in radius of the synchrotron loss is faster than the advection gain,  $\dot p_{syn}\propto r^{-5/2}$ versus $\dot p_{adv}\propto r^{-3/2}$, the advection dominated region becomes smaller and smaller and disappears for radii very close to the BH horizon~\cite{Regis:2008ij}. As stated above, in the region with $r > R_{acc}$ we neglect the advection and thus the trajectories are just horizontal lines.

A convection effect, due to the outflow of primary cosmic rays from the Galactic disk, where sources are located, could be relevant also for the propagation of  particles in the whole Milky Way diffusion 
region~\cite{Strong:2001gh}; wind velocities are expected however to be much smaller and their effect much less dramatic.

\subsection{Multi-frequency Emissivities and Intensities}\label{sec:emissions}

The solution of Eq.~\ref{eq:prop_n} provides the $e^+/e^-$ number density $n_e$ in the 
stationary limit.  For a radiative process $i$, with associated power $P_{i}$, the photon 
emissivity at frequency $\nu$ is obtained by folding $n_e$ with the power \cite{1979rpa..book.....R}:
\begin{equation}
j_{i}(\vec x, \nu)=2\int^{M_{\chi}}_{m_e}dE\, P_{i}(\vec x,E,\nu)\, n_e(\vec x,E)\;,
\label{eqjsynch}
\end{equation}
where the factor $2$ takes into account electrons and positrons. The corresponding intensity
$I_i$, as measured by a detector, follows from solution of the differential equation:
\begin{equation}
\frac{dI_{i}(\nu,s)}{ds}=-\alpha(\nu,s)\,I_i(\nu,s,\tilde \theta)+\frac{j_i(\nu,s)}{4\pi}\,,
\label{eqdiffIsynch}
\end{equation}
i.e. within the increment $ds$ along a line of sight of observation, there is a gain in intensity $j_i/(4\,\pi)\,ds$, while a decrease $\alpha\,I_i\,ds$ could be due to absorption.

\subsubsection{Radio emission}

At radio frequencies, the DM-induced  emission is dominated by the synchrotron radiation.
The power for synchrotron emission takes the form \cite{1979rpa..book.....R}:
\be
P_{syn} (\vec x,E,\nu)= \frac{\sqrt{3}\,e^3}{m_e c^2} \,B(\vec x) F(\nu/\nu_c)\;,
\ee
where $m_e$ is the electron mass, the critical synchrotron frequency is defined as $\nu_c \equiv  3/(4\,\pi) \cdot {c\,e}/{(m_e c^2)^3} B(\vec x) E^2$,  and $F(t) \equiv t \int_t^\infty dz K_{5/3}(z)$ is the standard function setting the spectral behavior of synchrotron radiation. Synchrotron self-absorption can be a relevant effect; indeed one finds that, when considering the region around the Galactic center, the synchrotron luminosity is so large that low frequency intensities are significantly changed by absorption~\cite{Gondolo:2000pn,Bertone:2001jv,Aloisio:2004hy}; in principle, synchrotron self-Compton effects
could be sizable as well, but they turn out to subdominant~\cite{Aloisio:2004hy}. For all other sources,
absorption can be safely neglected.

It is instructive to compare the synchrotron to the gamma-ray luminosity under a few simplifying assumptions. Consider the limit of radiative losses as dominant term in the transport equation,
so that the electron/positron equilibrium number density is given by Eq.~\ref{eq:nodiff}. Suppose
also that the energy loss rate is dominated by synchrotron emission, i.e.  $\dot p \simeq \dot p_{syn}$,
and that it is sufficient to implement the monochromatic approximation for the synchrotron power, i.e. assumed  $F(\nu/\nu_c)\sim \delta(\nu/\nu_c-0.29)$ \cite{1979rpa..book.....R}. In the monochromatic approximation there is a one-to-one correspondence between the energy of the radiating electron (peak energy in the power) and the frequency of the emitted photon, that is $E_p=\nu^{{1}/{2}}(0.29\, B \,c_0)^{-{1}/{2}}$ with $c_0=3/(4\,\pi) \cdot {c\,e}/{(m_e c^2)^3}$, or, introducing values for numerical constants, the peak energy in GeV is $\widehat{E}_{p}\simeq 0.463\,\widehat{\nu}^{{1}/{2}}\widehat{B}^{-{1}/{2}}$, with $\widehat{\nu}$ the frequency in ~MHz and $\widehat{B}$ the magnetic field in~$\mu$G. 
Under this set of approximations, the induced synchrotron luminosity becomes:
\begin{equation}
\nu L_{\nu}^{syn} =\,\frac{9\sqrt{3}}{4} \frac{\sigma v}{M_{\chi}^2}\int d^3x \rho(\vec x)^2  E_p \, Y_e(E_p)\,,
\label{eq:nuLnus}
\end{equation}
where we have defined $Y_e(E)=\int^{M_{\chi}}_E dE' dN_e/dE'$. Analogously, the induced 
$\gamma$--ray luminosity is:
\begin{equation}
\nu L_{\nu}^{\gamma}=\,2\pi\, \frac{\sigma v}{M_{\chi}^2}\int d^3x \rho(\vec x)^2\,E^2\,\frac{dN_{\gamma}}{dE} \;.
\label{eq:nuLnug}
\end{equation}
Having already stressed that electron/positron yield from WIMP annihilation is at the same level than 
the $\gamma$--ray yield, it follows that the radio and $\gamma$--ray luminosities are also expected
to be roughly comparable. 

\subsubsection{From the UV to the  gamma-ray band}

For large magnetic field, such as for the Galactic center case, synchrotron emission may extend to
the UV and (possibly) the X-ray band. The emission through Inverse Compton (IC) scattering of the ultra--relativistic electrons from WIMP annihilations on cosmic microwave and starlight background photons give rise to a spectrum of photons stretching from below the extreme
ultra-violet up to the soft gamma-ray band, peaking in the soft X-ray energy band.
The Inverse Compton power is given by:
\be
P_{IC}(\vec x,E,\nu) = c\,h\nu \int d\epsilon\, \frac{dn_\gamma}{d\epsilon}(\epsilon,\vec x)\,\sigma(\epsilon,\nu,E) 
 \label{eq:PIC}
\ee
where $\epsilon$ is the energy of the target photons, $dn_\gamma/d\epsilon$ is their differential energy
spectrum and  $\sigma$ is the Klein--Nishina cross section. Additional radiative emission is expected
from the process of non-thermal bremsstrahlung, {\em i.e.} the emission of gamma-ray photons in the deflection of the charged particles by gas electrostatic potential, as well as via ionization and Coulomb scattering; in general, these processes give subdominant contributions with respect to synchrotron and inverse Compton emission.

At gamma-ray frequencies, radiative emission matches the hard gamma-ray component which,
as already mentioned arises from prompt emission in WIMP pair annihilations. 
%
\begin{figure}[!t]
\begin{center}
\hspace*{-0.5cm}\includegraphics[width=8.5cm]{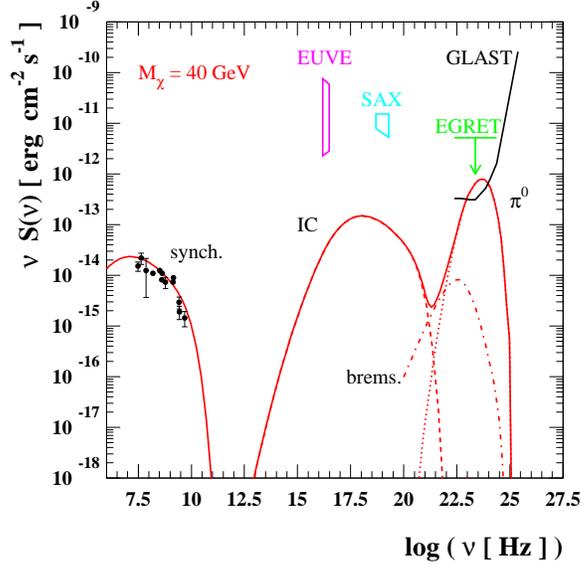}\\
\end{center}
\caption{The multi-wavelength spectrum from a 100 GeV WIMP annihilating in $b$$\bar b$ in the dwarf spheroidal Galaxy Draco. The WIMP pair annihilation rate has been tuned to give a gamma-ray signal at the level of the EGRET limit on the gamma-ray emission.
}
\label{fig:multiw}
\end{figure}
%
We present in fig.~\ref{fig:multiw} an example of the multi-wavelength spectrum from a 100 GeV WIMP annihilating in $b$$\bar b$ in the dwarf spheroidal Galaxy Draco, with the details of the origin of radiation at various frequencies (from Ref.~\cite{Colafrancesco:2006he}).

\section{The multi-wavelength approach and Galaxy Clusters}\label{sec:clusters}
Analyses of multi-wavelength signals from dark matter have been discussed in the context of galaxy clusters for a few case studies, including the Coma cluster \cite{Colafrancesco:2005ji}, the Bullet cluster \cite{Colafrancesco:2007gx} and of the Ophiuchus cluster \cite{Profumo:2008fy}. We give below a summary of the results of these studies.
\subsection{The case of Coma Cluster}\label{sec:coma}

Ref.~\cite{Colafrancesco:2005ji} performs a thorough analysis of the transport and diffusion properties of WIMP annihilation products in the Coma cluster of galaxies, investigating the resulting multi-frequency signals, from radio to gamma-ray frequencies. Other relevant astrophysical effects of WIMP
annihilations are also discussed, including the DM-induced Sunyaev-Zel'dovich effect and intra-cluster gas
heating. Ref.~\cite{Colafrancesco:2005ji} finds that Coma radio-halo data (the spectrum and the surface brightness) can be
nicely fitted by the WIMP-induced signal for certain particle physics models and
 under reasonable assumptions for the structure of the
intra-cluster magnetic field. Fitting the radio data and moving
to higher frequencies, \cite{Colafrancesco:2005ji} points out that the multi-frequency spectral energy distributions are
typically dim at EUV and X-ray frequencies (with respect to the data), but show a
non-negligible gamma-ray emission, depending on the amplitude of the Coma magnetic field.
A simultaneous fit to the radio, EUV and HXR data is not possible without violating the
gamma-ray EGRET upper limit. However, the inverse Compton emission can reproduce both the spectrum and the spatial distribution of the EUV emission observed in
Coma, provided that a quite small average magnetic field $B \sim 0.15$ \mug is assumed.
Such low value of the B field is also able to make the radio data and the hard X-ray data
of Coma consistent within a Synchrotron/IC model for their origins. The best-fit particle physics models yields substantial
heating of the intra-cluster gas, but not sufficient energy injection as to explain the
quenching of cooling flows in the innermost region of clusters. The best-fit ($b \bar{b}$) WIMP model also yields a detectable SZ effect (with a peculiar spectrum very
different from that of the thermal SZ effect) at the level of $\sim$ 40 to 10 $\mu$K in
the frequency range $\sim 10-200$ GHz, which could be observable with the next generation
high-sensitivity bolometric arrays, space and balloon-borne microwave experiments, like
PLANCK, OLIMPO, APEX, ALMA.

\subsection{The case of the Bullet Cluster}\label{sec:bul}

The SZ$_{DM}$ effect is an inevitable consequence of the presence and of the nature of DM
in large-scale structures. Ref.~\cite{Colafrancesco:2007gx} showed that microwave observations of the \es cluster can provide crucial probes for the presence and for the nature of DM in cosmic structures. The authors calculate the expected SZ effect from Dark Matter annihilation in the main mass
  concentrations of the cluster, and estimate the sources of contamination,
  confusion and bias to asses its significance. It is found that SZ observations at $\nu \approx 223$ GHz can resolve both spatially and
   spectrally the SZ$_{DM}$ signal and isolate it from the other SZ signals,
   and mainly from the thermal SZ effect which is null at $ \nu \sim 220-223$ GHz
   for the case of \es. SZ observations of the SZ$_{DM}$ effect with $\simlt$ arcmin resolution
   and $\simlt \mu$K sensitivity of \es are therefore crucial, and maybe unique, to find direct
   astrophysical probes of the existence and of the nature of Dark Matter,
   or to set strong experimental limits.

\subsection{The case of the Ophiuchus Cluster}\label{sec:oph}

Ref.~\cite{Profumo:2008fy} investigates a scenario where the recently discovered non-thermal hard X-ray emission from the Ophiuchus cluster \cite{Eckert:2007fv} originates from inverse Compton scattering of energetic electrons and positrons produced in WIMP dark matter pair annihilations. This scenario is compatible with all observational information on the cluster, and it can account for both the X-ray and the radio emission, provided the average magnetic field is of the order of \mbox{0.1 $\mu$G}. It is also shown that Fermi-LAT will conclusively test the dark matter annihilation hypothesis. Depending on the particle dark matter model, the LAT might even detect the monochromatic line produced by dark matter pair annihilation into two photons.

\section{The multi-wavelength approach and Dwarf Galaxies}\label{sec:dwarfs}
A class of low-background targets for multi-wavelength searches for WIMP dark matter annihilation is nearby dwarf galaxies. The general setup and an application to the specific case of the Draco dSph galaxy were discussed in \cite{Colafrancesco:2006he}, while constraints on WIMP models from XMM observations of the three satellite dSph galaxies Fornax, Carina and Ursa Minor have been discussed in \cite{2008ApJ...686.1045J}.

\subsection{The case of Draco}\label{sec:draco}
After investigating mass models for Draco in the light of
available observational data, Ref.~\cite{Colafrancesco:2006he} models the dark matter density profile, taking
advantage of numerical simulations of hierarchical structure formation. The gamma-ray and electron/positron yield expected for WIMP models is then analyzed. Unlike in larger dark matter structures -- such as galaxy
clusters -- spatial diffusion plays here an important role. While Draco would appear as a
point-like gamma-ray source, synchrotron emission from electrons and positrons produced
by WIMP annihilations features a spatially extended structure. Depending upon the cosmic
ray propagation setup and the size of the magnetic fields, the search for a diffuse radio
emission from Draco can be a more sensitive indirect dark matter search probe than gamma
rays. In some cases, an extended radio emission could be
detectable from Draco even if no gamma-ray source is identified by Fermi-LAT or by ACTs,
making this technique the most promising search for dark matter signatures from the class of
objects under consideration, i.e. nearby dwarf spheroidal galaxies. Ref.~\cite{Colafrancesco:2006he} also points out that available data are consistent with the presence of a black
hole at the center of Draco: if this is indeed the case, very significant enhancements of
the rates for gamma rays and other emissions related to dark matter annihilations are
expected.

\subsection{X-ray constraints from local Dwarfs}\label{sec:xraydwarf}

Local dSph galaxies are an ideal environment for particle dark matter searches with X-rays. Ref.~\cite{2008ApJ...686.1045J} describes how X-rays are produced as secondary radiation in Inverse Compton scattering off cosmic microwave background photons of electrons and positrons resulting from particle dark matter annihilation. The resulting spectrum is only mildly dependent on the details of the particle dark matter model (the dark matter mass and the dominant final state into which it pair annihilates), and it is, generically, hard (spectral index smaller than $\sim 1.5$). The normalization of the emission depends on (1) the particle dark matter pair annihilation rate, (2) the diffusion setup, and (3) the dark matter density distribution. For reasonable choices of these three {\em a priori} unknown inputs of the problem, the X-ray emission is potentially within reach of current X-ray detectors. Interestingly enough, the shape of the spectral energy distribution indicates that for dSph galaxies X-rays have a comparable, if not better, sensitivity to indirect dark matter detection than gamma rays.

Ref.~\cite{2008ApJ...686.1045J} used XMM-Newton archival data on three Local Group dSph galaxies, Ursa Minor, Fornax and Carina, to search for the diffuse X-ray emission expected from dark matter annihilation. Ref.~\cite{2008ApJ...686.1045J} studies the optimal energy and radial range to search for this type of emission, and concluded that for XMM-Newton and for the dSph galaxies under investigation these correspond to an energy band between 0.5 and 8 keV and to a radius of around $6^\prime$. No significant signal over background is found, and this, in turn, is turned into constraints on particle dark matter models. The best constraints result from both the Fornax and the Ursa Minor observations, while data from Carina result in bounds that are a factor of a few weaker.  Ursa Minor features the largest dark matter density, making it the best candidate target, but has the shortest usable XMM exposure.

In determining the impact on particle dark matter searches of our X-ray constraints, ref.~\cite{2008ApJ...686.1045J} points out the uncertainties resulting from the modeling of cosmic ray diffusion processes, and from the dark matter distribution. In particular, including dark matter substructures can boost our constraints significantly. The bounds obtained are expressed in terms of the dominant dark matter annihilation final states. For those final states relevant for specific dark matter models, such as supersymmetry, the constraints on the mass versus pair annihilation plane are very similar. In the most conservative setup only models with rather large annihilation cross sections are excluded. Assuming a smaller diffusion coefficient, or factoring in the effect of dark matter substructures, our constraints fall well within the interesting region where the supersymmetric dark matter can be a thermal relic from the early universe. Also, limits can be set on particular supersymmetric dark matter scenarios, such as Wino or Higgsino lightest neutralino dark matter.

An important result of the analysis is that even assuming a conservative diffusion setup the sensitivity of X-rays and of gamma rays to particle dark matter annihilation in dSph galaxies are comparable. This fact has two-fold implications: on the one hand, if longer observations of dSph galaxies were carried out with existing telescopes, it is possible that the first astronomical signature of particle dark matter annihilation would come from X-rays. Secondly, should a signature be detected in the future with gamma-ray telescopes, it would be extremely important to confirm the nature of the signal via X-ray observations.

Future X-ray telescopes, like Constellation-X and XEUS, will also have greatly increased effective areas with respect to current instruments. Using the currently available projections for the effective areas and backgrounds of these telescopes it is estimated that the X-ray limits (0.5-8 keV band) placed by a 100 ksec observation of Ursa Minor with Constellation-X or XEUS would improve over the limits placed in this paper by factors of roughly 35 and 70, respectively.  Thus even for a conservative diffusion model, the future generation of X-ray telescopes will place similar constraints on dark matter annihilation from dwarfs as Fermi-LAT, stronger constraints at particle masses below a few hundred GeV. In addition, a signal from Fermi could be confirmed with X-ray observations. 

In short, X-rays can play an important role in exploring the nature of particle dark matter and in pinpointing its properties. This role is complementary, but not subsidiary, to searches with gamma rays. Very exciting results at both frequencies might be just around the corner.

\section{The multi-wavelength approach and the Milky Way}\label{sec:mw}
Galactic sources, including the central regions of the Galaxy as well as galactic DM clumps, are interesting targets for multi-wavelength searches for WIMP annihilations. We summarize here results on the innermost galactic center region from Ref.~\cite{Regis:2008ij}, dark matter ``mini-halos'' \cite{Baltz:2004bb} and the dark matter interpretation of the so-called WMAP ``haze'' \cite{Hooper:2007kb}.

\subsection{The Galactic Center}\label{sec:gc}

Ref.~\cite{Regis:2008ij} presents a systematic, self-consistent study of the multi--wavelength emission due to WIMP pair annihilations in the Galactic center region. High-energy electrons and positrons emitted in a region with large magnetic fields give rise to synchrotron emission covering radio frequencies up to, possibly, the X-ray band. The authors discuss spectral and angular features,  and  sketch the correlations among signals in the different energy bands. The critical assumptions for this region stem from uncertainties in the DM source functions, regarding both WIMP models and DM distributions, and from the modeling of propagation for electrons and positrons, and the assumptions on magnetic field profiles. Radio to mm synchrotron emission is essentially independent from the shape of the magnetic field in the innermost region of the Galaxy, while at shorter wavelengths, i.e. in the infrared and, especially, the X--ray band, a different choice for the magnetic field may change predictions dramatically. Radio signals have in general very large angular sizes, larger than the typical size for the source function and hence of the $\gamma$-ray signals. The size of the region of synchrotron X-ray emissivity shrinks dramatically going to larger frequencies, smaller WIMP masses, or softer annihilation channels.  

The luminosity of the WIMP source at the different frequencies, and especially comparing the radio to the $\gamma$-ray band, is essentially at a comparable level, with luminosity ratios depending rather weakly on WIMP mass and annihilation channels. This is interesting, since the GC astrophysical source Sgr~A$^*$, an unusual source, certainly very different from typical galactic or extragalactic compact sources associated to black holes, has a very low luminosity over the whole spectrum, at a level at which it is plausible that a WIMP-induced component may be relevant. Indeed, while none of the fluxes detected in GC direction has spectral or angular features typical of a DM source, still all data-sets contribute to place significant constraints on the WIMP parameter space. Although the $\gamma$-ray band is the regime in which it is most straightforward to make the connection between a given dark matter model and the induced signal (hence it is also the regime on which most of previous analyses have concentrated on),  it does not seem to be the energy range with the best signal to background ratios. In the case of large magnetic fields close to the GC, X-ray data can give much tighter constraints. Radio and NIR measurements, which are less model dependent, tend to be more constraining as well.

\subsection{Galactic Mini-halos}\label{sec:minihalos}

Ref.~\cite{Baltz:2004bb} calculates the spectrum and spatial extent of diffuse emission from the charged particle products of
dark matter annihilations in galactic satellites that are currently within the
diffusion zone, namely within a few kpc of the stellar disk. For satellites moving with typical galactic halo velocities
of 300 km s¡Ý1 , the crossing time of the diffusion zone is of the same order as the diffusion time, thus an inherently
time-dependent treatment is required. For annihilation sources, e.g. galactic satellites at typical distances of 10 kpc, the diffuse emission in both inverse
Compton and synchrotron extends over roughly 300 square degrees. At least in terms of the
number of photons, the diffuse inverse Compton emission might be detectable by Fermi-LAT, assuming bright enough
annihilation sources. The spatial extent of the emission makes however its detection problematic.

\subsection{The WMAP ``Haze''}\label{sec:haze}

Ref.~\cite{Hooper:2007kb} shows that the observed features of the WMAP haze match the expected signal produced through the synchrotron
emission of dark matter annihilation products for a model with a cusped halo profile scaling as $\rho(r) \propto r^{-1.2}$ in the inner
kpc, and an annihilation cross section of $\sim 3 \times 10^{-26}$ cm$^3$/s. A wide range of annihilation modes are consistent with the synchrotron spectrum, and no boost factors are required. The properties required of a WIMP to generate the haze are precisely those anticipated for the most theoretically attractive particle dark matter candidates. If the haze is generated through dark matter annihilations, this will have very interesting implications for Fermi. If the $\rho(r) \propto r^{-1.2}$ slope of the halo profile continues to the inner Galaxy, the gamma ray flux from the Galactic Center will be observable by Fermi, so long as the WIMP is lighter than several hundred GeV, in spite of the presence of the observed HESS source in the region. 

\section{Radio Observations}\label{sec:radio}

We summarize here the findings of two studies, \cite{Borriello:2008gy} and \cite{Zhang:2008rs}, that addressed the detection of a radio emission from WIMP annihilation.

Ref.~ \cite{Borriello:2008gy} calculates the secondary WIMP-induced
radio emission from the galactic halo as well as from its expected
substructures, comparing it with the measured diffuse radio
background. They employ a multi-frequency approach using data in the
relevant frequency range 100 MHz--100 GHz, as well as the WMAP
Haze data at 23 GHz. The derived constraints are of the order
$\sigva= 10^{-24}$\,cm$^3$s$^{-1}$ for a DM mass
$m_{\chi}=100$\,GeV sensibly depending however on the
astrophysical uncertainties, in particular on the assumption on
the galactic magnetic field model. The
signal from single bright clumps offers only weak signals
because of diffusion effects which spread the electrons over large
areas diluting the radio signal. The diffuse signal from the halo
and the unresolved clumps is instead  relevant and can be compared
to the radio astrophysical background to derive constraints on the
DM mass and annihilation cross section. Conservative constraints are
at the level of $\sigva \sim 10^{-23}$\,cm$^3$s$^{-1}$ for a DM mass
$m_{\chi}=100$\,GeV from the WMAP Haze at 23 GHz. However, depending
on the astrophysical uncertainties, in particular on the assumption
on the galactic magnetic field model, constraints as strong as
$\sigva \sim 10^{-25}$\,cm$^3$s$^{-1}$ can be achieved.

Ref.~\cite{Zhang:2008rs} calculates the intensity and angular power spectrum of the cosmological background of
synchrotron emission from WIMP annihilations into electron positron pairs, and compares this background
with intensity and anisotropy of astrophysical and cosmological radio backgrounds, such
as from normal galaxies, radio-galaxies, galaxy cluster accretion shocks, the cosmic
microwave background and with Galactic foregrounds.
Under modest assumptions for the dark matter clustering, Ref.~ \cite{Zhang:2008rs} finds that around 2 GHz
average intensity and
fluctuations of the radio background at sub-degree scales allows to probe dark matter
masses $\ga100\,$GeV and annihilation cross sections not far from the natural values
$\left\langle\sigma v\right\rangle\sim3\times10^{-26}\,{\rm cm}^3\,{\rm s}^{-1}$. The angular power spectrum
of the signal from dark matter annihilation tends to be flatter than that from astrophysical
radio backgrounds. Furthermore, radio source counts have comparable constraining power.
Such signatures are interesting especially for future radio detectors such as SKA.

\section{Conclusions and Overview}\label{sec:concl}
If dark matter particles pair annihilate into stable Standard Model particles, a population of energetic, non-thermal electrons and positrons is necessarily produced. The secondary radiation resulting from the energy losses of this non-standard population (including synchrotron and inverse Compton up-scattering of background radiation photons) in turn produces a peculiar multi-wavelength spectrum, extending from radio to gamma-ray frequencies. In some cases, constraints from radio or X-ray observations of galactic or extragalactic targets place constraints on dark matter models which are more stringent than those resulting from gamma-ray data. Reversing this statement, multi-wavelength observations will complement gamma-ray observations as probes of particle dark matter, since the expected luminosities at different frequencies are generically comparable. The indirect search for dark matter with astronomical observations at various frequencies is therefore a crucial and indispensable element in the quest for the fundamental nature of dark matter.


\bibliographystyle{plainyr}

\begin{thebibliography}{10}

\bibitem{Melia:1992}
F.~Melia, M.~Wardle, P.~Heimberg, M.~Wardle, and P.~Heimberg.
\newblock {On the stability of neutron star winds}.
\newblock 1992.
\newblock STEWARD-1035.

\bibitem{Jungman:1995df}
Gerard Jungman, Marc Kamionkowski, and Kim Griest.
\newblock {Supersymmetric dark matter}.
\newblock {\em Phys. Rept.}, 267:195--373, 1996.

\bibitem{Colafrancesco:1998us}
Sergio Colafrancesco and Pasquale Blasi.
\newblock {Clusters of galaxies and the diffuse gamma-ray background}.
\newblock {\em Astropart. Phys.}, 9:227--246, 1998.

\bibitem{Baltz:1998xv}
Edward~A. Baltz and Joakim Edsjo.
\newblock {Positron propagation and fluxes from neutralino annihilation in the
  halo}.
\newblock {\em Phys. Rev.}, D59:023511, 1999.

\bibitem{Blasi:1999aj}
Pasquale Blasi and Sergio Colafrancesco.
\newblock {Cosmic rays, radio halos and nonthermal X-ray emission in clusters
  of galaxies}.
\newblock {\em Astropart. Phys.}, 122:169--183, 1999.

\bibitem{Gondolo:2000pn}
Paolo Gondolo.
\newblock {Either neutralino dark matter or cuspy dark halos}.
\newblock {\em Phys. Lett.}, B494:181--186, 2000.

\bibitem{Bertone:2001jv}
G.~Bertone, G.~Sigl, and J.~Silk.
\newblock {Astrophysical limits on massive dark matter}.
\newblock {\em Mon. Not. Roy. Astron. Soc.}, 326:799--804, 2001.

\bibitem{Strong:2001gh}
A.~W. Strong and I.~V. Moskalenko.
\newblock {New developments in the GALPROP CR propagation model}.
\newblock 2001.

\bibitem{Aloisio:2004hy}
Roberto Aloisio, Pasquale Blasi, and Angela~V. Olinto.
\newblock {Neutralino annihilation at the galactic center revisited}.
\newblock {\em JCAP}, 0405:007, 2004.

\bibitem{Baltz:2004bb}
Edward~A. Baltz and L.~Wai.
\newblock {Diffuse inverse Compton and synchrotron emission from dark matter
  annihilations in galactic satellites}.
\newblock {\em Phys. Rev.}, D70:023512, 2004.

\bibitem{Colafrancesco:2005ji}
Sergio Colafrancesco, S.~Profumo, and P.~Ullio.
\newblock {Multi-frequency analysis of neutralino dark matter annihilations in
  the Coma cluster}.
\newblock {\em Astron. Astrophys.}, 455:21, 2006.

\bibitem{Colafrancesco:2007gx}
Sergio Colafrancesco, P.~de~Bernardis, S.~Masi, G.~Polenta, and P.~Ullio.
\newblock {Direct probes of Dark Matter in the cluster 1ES0657-556 through
  microwave observations}.
\newblock 2007.

\bibitem{Colafrancesco:2006he}
Sergio Colafrancesco, S.~Profumo, and P.~Ullio.
\newblock {Detecting dark matter WIMPs in the Draco dwarf: a multi- wavelength
  perspective}.
\newblock {\em Phys. Rev.}, D75:023513, 2007.

\bibitem{Eckert:2007fv}
D.~Eckert, N.~Produit, S.~Paltani, A.~Neronov, and T.~J.~L. Courvoisier.
\newblock {INTEGRAL discovery of non-thermal hard X-ray emission from the
  Ophiuchus cluster}.
\newblock 2007.

\bibitem{Hooper:2007kb}
Dan Hooper, Douglas~P. Finkbeiner, and Gregory Dobler.
\newblock {Evidence Of Dark Matter Annihilations In The WMAP Haze}.
\newblock {\em Phys. Rev.}, D76:083012, 2007.

\bibitem{Hooper:2007qk}
Dan Hooper and Stefano Profumo.
\newblock {Dark matter and collider phenomenology of universal extra
  dimensions}.
\newblock {\em Phys. Rept.}, 453:29--115, 2007.

\bibitem{Picozza:2006nm}
P.~Picozza et~al.
\newblock {PAMELA: A payload for antimatter matter exploration and light-nuclei
  astrophysics}.
\newblock {\em Astropart. Phys.}, 27:296--315, 2007.

\bibitem{Strong:2007nh}
Andrew~W. Strong, Igor~V. Moskalenko, and Vladimir~S. Ptuskin.
\newblock {Cosmic-ray propagation and interactions in the Galaxy}.
\newblock {\em Ann. Rev. Nucl. Part. Sci.}, 57:285--327, 2007.

\bibitem{Baltz:2008wd}
E.~A. Baltz et~al.
\newblock {Pre-launch estimates for GLAST sensitivity to Dark Matter
  annihilation signals}.
\newblock {\em JCAP}, 0807:013, 2008.

\bibitem{Borriello:2008gy}
Enrico Borriello, Alessandro Cuoco, and Gennaro Miele.
\newblock {Radio constraints on dark matter annihilation in the galactic halo
  and its substructures}.
\newblock 2008.

\bibitem{2008ApJ...686.1045J}
T.~E. {Jeltema} and S.~{Profumo}.
\newblock {Searching for Dark Matter with X-Ray Observations of Local Dwarf
  Galaxies}.
\newblock {\em \apj}, 686:1045--1055, October 2008.

\bibitem{Komatsu:2008hk}
E.~Komatsu et~al.
\newblock {Five-Year Wilkinson Microwave Anisotropy Probe (WMAP)
  Observations:Cosmological Interpretation}.
\newblock 2008.

\bibitem{Porter:2008ve}
Troy~A. Porter, Igor~V. Moskalenko, Andrew~W. Strong, Elena Orlando, and
  Laurent Bouchet.
\newblock {Inverse Compton Origin of the Hard X-Ray and Soft Gamma- Ray
  Emission from the Galactic Ridge}.
\newblock {\em Astrophys. J.}, 682:400--407, 2008.

\bibitem{Profumo:2008fy}
Stefano Profumo.
\newblock {Non-thermal X-rays from the Ophiuchus cluster and dark matter
  annihilation}.
\newblock {\em Phys. Rev.}, D77:103510, 2008.

\bibitem{Regis:2008ij}
Marco Regis and Piero Ullio.
\newblock {Multi-wavelength signals of dark matter annihilations at the
  Galactic center}.
\newblock {\em Phys. Rev.}, D78:043505, 2008.

\bibitem{Zhang:2008rs}
Le~Zhang and Guenter Sigl.
\newblock {Dark Matter Signatures in the Anisotropic Radio Sky}.
\newblock {\em JCAP}, 0809:027, 2008.

\bibitem{1979rpa..book.....R}
G.~B. {Rybicki} and A.~P. {Lightman}.
\newblock {\em {Radiative processes in astrophysics}}.
\newblock New York, Wiley-Interscience, 1979.~393 p., 1979.

\end{thebibliography}

\end{document}